\documentstyle[12pt]{article}
\begin{document}

\def\lsim{\mathrel{\rlap{\lower3pt\hbox{\hskip0pt$\sim$}}
    \raise1pt\hbox{$<$}}}         
\def\gsim{\mathrel{\rlap{\lower4pt\hbox{\hskip1pt$\sim$}}
    \raise1pt\hbox{$>$}}}         
\def\dblint{\mathop{\rlap{\hbox{$\displaystyle\!\int\!\!\!\!\!\int$}
}
    \hbox{$\bigcirc$}}}
\def\ut#1{$\underline{\smash{\vphantom{y}\hbox{#1}}}$}
 
\newcommand{\beq}{\begin{equation}}
\newcommand{\eeq}{\end{equation}}
\newcommand{\aver}[1]{\langle #1\rangle}

\newcommand{\La}{\overline{\Lambda}}
\newcommand{\Lam}{\Lambda_{QCD}}

\newcommand{\ind}[1]{_{\begin{small}\mbox{#1}\end{small}}}
\newcommand{\hscale}{\mu\ind{hadr}}

\newcommand{\appa}{\mbox{\ae}}
\newcommand{\CP}{{\em CP } }
\newcommand{\fy}{\varphi}
\newcommand{\hi}{\chi}
\newcommand{\al}{\alpha}
\newcommand{\as}{\alpha_s}
\newcommand{\gf}{\gamma_5}
\newcommand{\gm}{\gamma_\mu}
\newcommand{\gn}{\gamma_\nu}
\newcommand{\be}{\beta}
\newcommand{\ga}{\gamma}
\newcommand{\de}{\delta}
\renewcommand{\Im}{\mbox{Im}\,}
\renewcommand{\Re}{\mbox{Re}\,}
\newcommand{\GeV}{\,\mbox{GeV}}
\newcommand{\MeV}{\,\mbox{MeV}}
\newcommand{\matel}[3]{\langle #1|#2|#3\rangle}
\newcommand{\state}[1]{|#1\rangle}
\newcommand{\ra}{\rightarrow}
\newcommand{\ve}[1]{\vec{\bf #1}}

\newcommand{\eq}[1]{eq.\hspace*{.1em}(\ref{#1}) }
 
\newcommand{\re}[1]{Ref.~\cite{#1}}
\newcommand{\res}[1]{Refs.~\cite{#1}}
\begin{titlepage}
\renewcommand{\thefootnote}{\fnsymbol{footnote}}
\begin{center} 

{\Large \bf Theoretical Physics Institute\\
University of Minnesota}
\end{center}
\begin{flushright}
TPI-MINN-94/13-T\\
UMN-TH-1251-94\\
NSF-ITP-94-39\\
hep-ph/9405207\\
\end{flushright}
\vspace{.4cm}
\begin{center} 
{\large\bf $|V_{cb}|$ from OPE Sum Rules for Heavy Flavor 
Transitions}
\end{center}
\vspace*{.5cm}
\begin{center} 

{\bf M. Shifman, N.G. Uraltsev\footnote{Permanent address: Theory 
Division, 
Petersburg Nuclear Physics Institute, Gatchina, St.Petersburg, 188350 
Russia} 
and
A. Vainshtein\footnote{also at Budker Institute of Nuclear Physics, 
Novosibirsk
630090, Russia}}\\ 
\vspace*{.4cm}
{\it  Theoretical Physics Institute, University of Minnesota,
Minneapolis, MN 55455}\\
\vspace*{.4cm}
e-mail addresses:\\
{\normalsize\it  SHIFMAN@VX.CIS.UMN.EDU,
VAINSHTE@VX.CIS.UMN.EDU}
\vspace*{2cm}

{\Large\bf Abstract}
\end{center}
\vspace*{.4cm}

We derive a  model-independent upper bound for the axial-vector 
form factor of the 
$B\ra D^*$ transition at zero recoil, $F_{B\ra D^*}$. The form factor 
turns out to be 
noticeably less than unity. The deviation of $F_{B\ra D^*}$ from  
unity is larger than previously anticipated. Using our
estimate we extract $|V_{cb}|$ from the measured exclusive rate
of $B\ra D^* l\nu$ extrapolated to the point of zero recoil. The 
central `` exclusive"
value of $|V_{cb}|$ is in agreement with the value obtained 
from the inclusive semileptonic width $\Gamma (B\ra X_c l\nu)$.
We argue that the  
theoretical uncertainty in  determining  $|V_{cb}|$
from the total inclusive width is significantly reduced
if a constraint on the quark mass difference $m_b - m_c$
stemming from the heavy quark expansion is taken into account.

\end{titlepage}
\addtocounter{footnote}{-2}

1. Precision determination of $|V_{cb}|$, the CKM matrix element, is 
one
of the most important practical applications of the heavy quark 
expansions 
in 
the $B$ physics today. Two methods 
allowing one to extract $|V_{cb}|$ from  data are commonly used:
the inclusive approach (from the total semileptonic width of the
$B$ meson) and the
exclusive one (based on the decay amplitude $B\ra D^* e\nu$ 
extrapolated 
to the point of zero recoil). Both methods have their own advantages 
and
drawbacks, purely experimental and theoretical. 
In this letter we address the problem of the theoretical uncertainties
unavoidable in obtaining $|V_{cb}|$ from experimental
data given the status of the present-day QCD.
The theoretical uncertainty usually quoted for
$|V_{cb}|$ dominates all other error bars, see e.g. \cite{Vcb}.
Our task is to show that it can be significantly
reduced provided that full information stemming from the 
fundamental QCD
is properly used. Both, the ``exclusive" and ``inclusive"
values of $|V_{cb}|$ will be considered. The exclusive method has
a larger experimental error due to lower statistics and the need of
extrapolation to the point of zero recoil. The measurements
of the inclusive semileptonic rates are more accurate.

On the theoretical side, the  
sources of  uncertainty in the
two approaches above are different. In the inclusive method
the theoretical expression for the width depends on the 
$b$ and $c$ quark masses which are allowed to vary independently 
within 
certain limits. In the exclusive method the amplitude is expressed 
directly 
in terms of known masses of $B$ and $D^{*}$ mesons; however, 
it is the $B\ra D^*$ form factor at zero recoil, $F_{B\ra D^*}$,
that is not known exactly. According to the theorem of \re{VS}
(see also \cite{nussinov})
in the limit $m_{b,c}\ra\infty$ ($m_b/m_c$ fixed) 
the appropriately normalized
form factor $F_{B\ra D^{*}}$ is equal to unity. 
For the actual values of the
quark masses there exist corrections in the inverse powers of
the masses. Linear corrections are absent at zero recoil 
\cite{VS,luke},
and the leading nonperturbative ones are quadratic in $1/m_{b,c}$.  
So far, they have not been 
calculated in a model-independent way although some estimates
exist in the literature.

We go beyond  the theorem of Refs. \cite{VS,luke} presenting a 
model-independent bound on the size of $1/m_{b,c}^2$ corrections
in terms of  the expectation value of
the chromomagnetic operator, see eq. (\ref{mug}). The value of 
$F_{B\ra D^*}$ is certainly less than 
unity. The bound we get indicates that the 
deviation of $F_{B\ra D^*}$ from unity (see eq. (\ref{prefer}$a,b$) 
below)  is beyond  previous estimates. 

Moreover, under some very plausible additional assumptions
we are able to convert the bound in an estimate of the actual value of 
$F_{B\ra D^*}$, see eq.  (\ref{prefer}$c$).

(The size of the $1/m_{b,c}^2$  corrections to
$F_{B\ra D^*}$ at zero recoil has
been previously discussed in \re{FN} with
the conclusion that the absolute value of this deviation is 
very small, less
than $0.03$ \footnote{
A. Falk suggested, however \cite{Falk}, that the estimate of Ref. 
\cite{FN},
$|1-F_{B\ra D^*}|=\pm 0.03$, should be rather perceived
as a ``$1\sigma$ error bar", and doubling the estimate
to achieve the ``$2\sigma$ confidence level" is welcome.}. Another 
analysis
of the $1/m_{b,c}^2$ terms published recently \cite{mannel}
also demonstrates that $F_{B\ra D^*}<1$.)

The  CLEO result \cite{Vcb}
obtained by a linear extrapolation of 
the  $B\ra D^*$ distribution to the point of 
zero recoil is
\beq
F_{B\ra D^*}|V_{cb}| = 0.038\pm 0.006\; .
\label{exp}
\eeq
Since we use the experimental number for the purpose of illustration only
we do not quote the results of other extrapolations as well as 
the ARGUS numbers \cite{argus}.
The standard practice in analyzing
the exclusive experimental data is  as follows: 
in extracting $|V_{cb}|$ one uses the
``reference" value, $F_{B\ra D^*}=1$; then the numbers in eq. (\ref{exp})
can be read as the results for $|V_{cb}|$. The theoretical
uncertainty in $F_{B\ra D^*}$ is then added to the experimental error bars
quoted for $|V_{cb}|$.

If we use the exact
bound mentioned above as the actual value of  $F_{B\ra D^*}$
the result for $|V_{cb}|$ increases by 6\%, at least; 
the actual increase is  expected to be even larger, from 8 to 14\%.

Next we turn to the inclusive method and analyze the uncertainty
in the theoretical expression for the total inclusive
decay rate in the $b\ra c$ transition. Superficially,
the decay rate is proportional to $m_b^5$ (the
quark masses will be denoted by small $m$ with the corresponding
subscript), and even
a modest error in the $b$ quark mass, say, 100 MeV,
is translated in the $\pm 5 \%$ error in the numerical value 
of $|V_{cb}|$, coming on top of other theoretical
uncertainties. The total theoretical uncertainty is usually believed
to lie in the 10\% range. We observe that a large fraction of the
events is kinematically close to the small velocity (SV) limit
\cite{VS}. In the SV limit the inclusive decay rate does not depend
on $m_b$ and $m_c$ individually, but rather on the mass difference
$m_b-m_c$ that is known to a much better accuracy (see eq. 
(\ref{mdif})).
Although in the actual decays not all events occur in the SV regime
the total inclusive probability is only weakly sensitive to
$m_b+m_c$. Using the constraint on $m_b-m_c$ we extract
the ``inclusive" value of $|V_{bc}|$ with the theoretical uncertainty
close to $\pm 5\%$. The dominant part of this uncertainty comes
from our rather poor knowledge of $\mu_\pi^2$, the
expectation value of the kinetic energy operator, see eq. (\ref{mug}). 
The latter
is measurable, in principle, in the very same semileptonic
transitions.

\vspace*{.2cm}

2. The prediction for  $F_{B\ra D^*}$ stems from the OPE/HQET sum 
rules
derived at the point of zero recoil. 
Details of the derivation will be presented
elsewhere \cite{optical}. Here let us sketch main points.

The transitions we are interested in are $B\ra D^*$ and $B\ra$ vector
excitations. These transitions are generated by the axial-vector 
current, $A_\mu =\bar b\gamma_\mu\gamma_5c$. If the
momentum carried by the lepton pair is denoted by $q$, the zero 
recoil
point is achieved if $\vec q =0$ and $q_0 =\Delta M$ where
$\Delta M \equiv M_B-M_{D^*}$. To obtain the sum
rule we consider the $T$ product
\beq
h_{\mu\nu} = i\int d^4x
{\rm e}^{-iqx}\frac{1}{2M_{H_Q}}
\langle B |T\{ A_\mu^\dagger (x) A_\nu (0)\}|B\rangle 
\eeq
assuming that $\vec q =0$ and $q_0$ is close to $\Delta M$.
The hadronic tensor $h_{\mu\nu}$ can be systematically expanded
in $\Lam /m_{b,c}$. For our purposes
it is sufficient to keep the terms quadratic in this parameter.
In general the hadronic tensor $h_{\mu\nu}$ is decomposed
in the sum of five terms \cite{CGG}; in this way
five structure functions, $h_1$ to $h_5$ are introduced. In the zero 
recoil 
point only two independent structures survive.
For the spatial components of the axial-vector current we need to 
consider
only $h_1$ (we will systematically use the notations of Ref. 
\cite{Koyrakh}.)

The ${\cal O}(1/m_{b,c}^2)$ terms in $h_{\mu\nu}$ were calculated
in Refs. \cite{Koyrakh,manohar}. It is convenient  to introduce
\beq
\epsilon = M_B-M_{D^*}- q_0 =\Delta M - q_0\;\;,
\label{BD}
\eeq
Using eq. (A.1) of \re{Koyrakh} one can write down the result
in the form
\beq
-h_1 =\frac{1}{\epsilon}
-\frac{\mu_G^2-\mu_\pi^2}{2m_b}
\left( \frac{1}{3} -\frac{m_c}{m_b}\right)
\frac{1}{\epsilon (2m_c +\epsilon )}
+\left[ \frac{4}{3}\mu_G^2
-(\mu_G^2-\mu_\pi^2)\frac{q_0}{m_b}\right]
\frac{1}{\epsilon^2 (2m_c+\epsilon )} ,
\label{h-one}
\eeq
where $\mu_G^2$ and $\mu_\pi^2$ parametrize the
matrix elements of the chromomagnetic and kinetic energy 
operators,
\beq
\mu_G^2 =\frac{1}{2M_{H_b}}\langle H_b|
\bar b\,\frac{i}{2}\sigma_{\mu\nu}G^{\mu\nu}\,b|H_b\rangle
\;\;;\;\;\; 
\mu_\pi^2 =\frac{1}{2M_{H_b}}\langle H_b|
\bar b\,(i\vec{D})^2 \,b|H_b\rangle\;\; .
\label{mug}
\eeq
Now we can expand in $\Lam /\epsilon$ and
in $\epsilon / m_{b,c}$ keeping only the
term linear in $1/\epsilon $, and compare
the theoretical expression obtained in this way with the hadronic
saturation of $h_{\mu\nu}$. 
In the derivation of our  sum rule, to the accepted accuracy,
one can neglect the difference between $\Delta M=M_B-M_D$ and
$\Delta m \equiv m_b - m_c$. The corresponding effect
is inversely proportional to the heavy quark mass
and affects our result only in a higher order in $\Lam
/m_{b,c}\,$. We find in this way
\beq
F_{B\ra D^*}^2 + \sum_{i=1,2,...}F_{B\ra excit}^2
=1 -\frac{1}{3}\frac{\mu_G^2}{m_c^2}
-\frac{\mu_\pi^2-\mu_G^2}{4}\left(
\frac{1}{m_c^2}+\frac{1}{m_b^2}+\frac{2}{3m_cm_b}
\right) ,
\label{SR!}
\eeq  
where the sum on the left-hand side runs over all excited states
with the appropriate quantum numbers,
and all form factors are taken at the point of zero recoil.
The form factor $B\ra D^*$ at zero recoil is defined as
$$
\langle B|A_\alpha|D^*\rangle =-
\sqrt{4M_BM_{D^*}}\;\;F_{B\ra D^*} 
\:D^*_{\alpha}\;\;\ ,
$$
where $D^*_{\alpha}$ is the polarization vector of $D^*$.

It is convinient to rewrite the eq. (\ref{SR!}) in the form
\beq
1-F_{B\ra D^*}^2 =\frac{1}{3}\frac{\mu_G^2}{m_c^2}
+\frac{\mu_\pi^2-\mu_G^2}{4}\left(
\frac{1}{m_c^2}+\frac{1}{m_b^2}+\frac{2}{3m_cm_b}
\right) + \sum_{i=1,2,...}F_{B\ra excit}^2 ~~.
\label{SR!!}
\eeq  
If terms ${\cal O}(\Lam^2/m_Q^2)$ are 
neglected then  higher states
cannot be excited at zero recoil -- only
the elastic $B\ra D^*$ transition survives in this approximation -- and 
we 
recover the prediction 
$$
F_{B\ra D^*}= 1 \,\,\, {\rm (zero\,\, recoil)}\;\;,
$$
the most well-known  consequence of the heavy quark (or the 
Isgur-Wise 
\cite{isgur}) symmetry -- a symmetry first observed in the point of 
zero 
recoil in Refs. \cite{VS,nussinov}.
 
At the level 
${\cal O}(\Lam^2/m_Q^2)$ excitation of higher states already takes 
place; all
transition form factors squared are proportional to $\Lam^2/m_Q^2$. 
Simultaneously
the form factor of the elastic transition shifts from unity by a similar 
amount. (These facts have a transparent physical interpretation, see 
\cite{optical}.)
 
It is crucial that
the contribution of the excited states in the sum rule (\ref{SR!!}) is 
strictly
positive.
Therefore we arrive at the following inequality:
\beq
1-F^2_{B\ra D^*} > 
\frac{1}{3}\frac{\mu_G^2}{m_c^2}
+ \frac{\mu_\pi^2-\mu_G^2}{4}\left(
\frac{1}{m_c^2}+\frac{1}{m_b^2}+\frac{2}{3m_cm_b}
\right) \;\; .
\label{ineqa}
\eeq

For completeness, we quote here a  similar zero recoil 
sum rule for the vector form 
factor of the $B\ra D$ transition $F_{B\ra D}$:
\beq
F_{B\ra D}^2 + \sum_{i=1,2,...}F_{B\ra excit}^2
= 1 
-\frac{\mu_\pi^2-\mu_G^2}{4}\left(
\frac{1}{m_c}-\frac{1}{m_b}
\right)^2\;\;\;.
\label{SRV}
\eeq
This transition is measurable (in principle)  at zero recoil in the $B\ra 
D+\tau\nu_\tau$ decays. To derive eq. (\ref{SRV}) one must consider 
the
zero-zero component of the hadronic tensor induced by the vector 
current
$V_\mu =\bar b\gamma_\mu c$.

\vspace*{.2cm}

3. Thus, we find that the lower bound for  the 
deviation of the elastic form factor 
$F_{B\ra D^*}$
from unity at zero recoil is determined, non-perturbatively,
by a  local contribution consisting of two terms. 
(The perturbative correction will be included below).
The first parameter, $\mu_G^2$,  is expressed via the mass 
difference
of $B^*$ and $B$,
\beq
\mu_G^2=\frac{3}{4}(M^2_{B^*}-M^2_B)\approx 0.35\GeV^2 .
\eeq
As far as $\mu_\pi^2$ is concerned, it was pointed out recently 
\cite{motion} that this parameter is bounded from below. 
An improved lower bound 
\beq
\mu_\pi^2 > \mu_G^2
\label{posit}
\eeq
was obtained recently \cite{voloshin}, using a quantum-mechanical 
argument similar to that of Ref. \cite{motion} and, later, within the 
sum rules 
themselves \cite{optical}. (Quantum-mechanically eq. (\ref{posit})
stems from the fact that the Pauli hamiltonian is positive-definite.)

Therefore, neglecting the second (positive) term 
in eq. (\ref{ineqa}) we arrive at a 
model-independent 
lower bound
\beq
1-F_{B\ra D^*}^{nonpert} >\frac{1}{8}\frac{M^2_{B^*}-M_B^2}{m_c^2} 
\approx 0.035 .
\label{modind}
\eeq
A stronger result is obtained if one relies on the 
estimate of $\mu_\pi^2$ from the QCD sum rules \cite{Braun} 
\footnote{ 
Earlier estimates are also available \cite{neubert,bagan}. Note that
the author of Ref. \cite{neubert} essentially revoked his result in a
later publication \cite{neubert2} where it is claimed
that the value of $\mu_\pi^2$ is dramatically smaller -- even smaller 
than
the lower bound (\ref{posit}) \cite{rumors}.}  :
\beq
\mu_\pi^2 =(0.54\pm 0.12)\GeV^2\;\;.
\label{bra}
\eeq
Then
\beq
1-F_{B\ra D^*}^{nonpert} > 0.05 \,\, {\rm to}\,\, 0.07 \, .
\label{QCDsr}
\eeq

On top of this non-perturbative correction it is necessary
to take into account radiative corrections due to the hard gluon 
exchange. 
In terms of the hadronic states these corrections correspond to the 
contribution of sufficiently heavy excited states 
where 
the perturbative calculation of inclusive transition rates is justified.
The one-loop correction
has been  found in \cite{VS},
\beq
\eta_A^{pert}=1 
+\frac{\alpha_s}{\pi}\left(\frac{m_b+m_c}{m_b-
m_c}\log{\frac{m_b}{m_c}}-
\frac{8}{3}\right)\simeq 0.975
\label{pertur}
\eeq
where we used
the one-loop value of $\Lam$ (for consistency
it is mandatory to use the one-loop
value of $\Lam$ in the one-loop calculations) and the subscript $A$
marks the axial-vector current.

The renormalization-group improvement (summation of the terms 
$\as^{n}\log^n{m_b/m_c}$ and $\as^{n}\log^{(n-1)}{m_b/m_c}$)
has been carried out in \re{neubert3}; it  leads to the result
$$
\eta_A^{pert}\approx 0.985\;\;.
$$
It is not quite clear, though,  whether one can trust this apparent 
reduction of the perturbative correction.  Indeed, for the actual 
values
of the quark masses
$\ln {m_b/m_c}\approx 1.3$ can hardly be considered as a large 
parameter. 
Therefore, the  exact calculation of the ${\cal O}(\as^2)$ term
makes much more sense than the summation of the leading and the 
next-to-leading logs, and will ensure a more reliable and accurate
prediction than the above renormalization-group improvement.
Note that  keeping 
only the $\log{m_b/m_c}$ term  in the one-loop calculation results in 
a dramatically wrong estimate (cf. eq. (\ref{pertur})).  For practical 
purposes 
we will
use for the gluon radiative correction to $F_{B\ra D^*}$ the factor 
$\eta_A^{pert} = 0.98\,$.

Accounting for the hard-gluon radiative corrections 
makes  the matrix element $\mu_G^2$ and $\mu_\pi^2$  
renormalization-scale dependent; in particular, the first one gets 
somewhat enhanced 
due to
the anomalous dimension of the chromomagnetic operator 
\cite{falk2}. 
These effects can be readily accounted for and do not produce a 
noticeable
change (for details see \re{optical}) even if the normalization point  
is chosen
at $\mu\sim 1\GeV$.

To get an idea of the contribution of the  inelastic channels 
one may accept that their joint effect  varies between 0 and 100\%
of the non-perturbative correction in eq. (\ref{SR!!}). Then assembling 
all these numbers together we finally arrive at
\beq
F_{B\ra D^*} = \left\{
\begin{array}{lr}
<0.94 \;\;\;\;\mbox{ model-independent bound,} & (a)\\
<0.92 \;\;\;\;
\mu_\pi^2=0.54 \GeV^2, \,\,\, & (b)\\
 0.89 \pm 0.03\;\;\;\; \mbox{ educated guess,}& (c)
\end{array}
\right.
\label{prefer}
\eeq
where the last entry assumes our educated guess
on the excited state contribution and the central value for the kinetic 
energy operator, eq. (\ref{bra}). 

To substantiate this estimate of the inelastic 
contribution let us consider a subclass of possible inelastic
contributions in the sum rule (\ref{SR!!}) due to the final states of the 
type $D\pi$ with $|{\vec p}_\pi |\ll \mu_{hadr}$. The soft-pion 
corrections in the elastic transition $B\ra D^*$ were studied 
previously in Ref. \cite{randall}. We follow the same pattern
but calculate, instead, the inelastic $D\pi$ contribution in the
sum rule.

For soft pions the amplitude $\langle D\pi |A_i|B\rangle $
is given by the diagrams of Fig. 1 and is reliably calculable,
\beq
\langle D^-\pi^+ |\vec A|B^+\rangle =
-\lambda \sqrt{4M_BM_D}\;{\vec p}_\pi\left(
\frac{1}{\epsilon}-\frac{1}{\epsilon+\Delta}\right)
\label{ampli}
\eeq
where $\epsilon$ is defined in eq. (\ref{BD}), ${\vec p}_\pi$
is the pion momentum,
$$
\Delta = M_{B^*}-M_B +M_{D^*} -M_D ,
$$
 and $\lambda$ is the heavy-meson-pion
constant,
$$
{\cal L}_{int} = 2M_D\,\lambda D^*_\mu D \partial^\mu\pi 
+2M_B\, \lambda B^*_\mu B\partial^\mu\pi\;\;,
$$
(for $D^{*0}D^-\pi^+$).  The $D^*D\pi$ and $B^*B\pi$ constants
are related by the heavy quark plus chiral symmetry
\cite{wiseb}; $1/m_{b,c}$ deviations from the heavy quark 
symmetry in
the vertices are inessential for the infrared part we are interested in.
In eq. (\ref{ampli}) we neglected also some other irrelevant terms
of higher order in $1/m_{b,c}$. Equation (\ref{ampli}) explicitly 
shows
cancellation of two graphs of Fig. 1 in the limit of the heavy
quark symmetry, when $\Delta \ra 0,\,\,\, \epsilon \gg\Delta$. This 
cancellation is just a 
manifestation of the fact that $F_{B\ra D^*}$ at zero recoil must be 
equal to unity up to $1/m_{b,c}^2$ corrections \cite{VS,luke}.
At $\epsilon <\Delta $ no trace of the heavy quark symmetry 
is left and no cancellation occurs; the diagram 1$a$ is much smaller
than that of Fig. 1$b$. The absence of the heavy quark symmetry in 
similar  kinematical conditions has been noted previously in Ref. 
\cite{dosch}.

The $D\pi$ production threshold is situated at
$$
\epsilon_0 = - (M_{D^*}-M_D-M_\pi ) ,
$$
i.e. at a small negative value of $\epsilon$. The pole at 
$\epsilon = 0$ in eq. (\ref{ampli}) evidently corresponds to
the actual production of $D^*$ with the
subsequent decay into $D\pi$. The second singularity,
the $B^*$ meson pole $(\epsilon +\Delta )^{-1}$, (see Fig. 1$a$)
lies outside the physical domain $\epsilon >\epsilon_0$.

The amplitude (\ref{ampli}) implies the
following hadronic tensor:
\beq
\frac{1}{\pi}{\rm Im}\, h_1|_{D\pi}
=\frac{\lambda^2}{8\pi^2}|{\vec p}_\pi |^3
\frac{\Delta^2}{\epsilon^2 (\epsilon +\Delta^2)^2}.
\label{hdpi}
\eeq
Here we also added the channel ${\bar D}^0\pi^0$. Integrating this 
expression 
over $\epsilon$ we get  the contribution to the sum rule (\ref{SR!})
sought for, cf. eq. (\ref{h-one}).

To make the integral
\beq
\frac{1}{\pi}\int d\epsilon \, {\rm Im}\, h_1
\label{inth}
\eeq
well-defined the factor $\epsilon^{-2}$ has to be replaced
by $(\epsilon^2 +(\Gamma^2/4))^{-1}$ where $\Gamma$ is the pion
width of $D^*$. Then integration near $\epsilon = 0$ yields
unity -- this is nothing else than the leading elastic contribution to the sum
rule (\ref{SR!}) which has to be unity in the calculation at hand.
It must be removed from our inelastic part. To this end the lower 
limit of integration in eq. (\ref{inth}) must be chosen at some
$\epsilon_{min}\gg \Gamma$.  The upper limit of integration
is also needed since the integral (\ref{inth}) is logarithmically
divergent at large $\epsilon$. This is a standard situation with the
soft-pion amplitudes containing chiral logarithms. We will
cut off the integral at $\epsilon =\mu_{hadr}\sim 1$ GeV, so that
the expression (\ref{ampli}) for the amplitude stays valid
inside the integration range. Notice that not only the
coefficient in front of the chiral logarithm is reliably calculable;
the constant term from the domain $\epsilon\sim \Delta$
also comes out correctly.

Doing the integral we find that
\beq
\sum F^2_{B\ra excit} \ra
\frac{1}{\pi}\int d\epsilon \, {\rm Im}\, h_1|_{D\pi}=
\frac{\lambda^2\Delta^2}{8\pi^2}\left( \ln\frac{\mu_{hadr}}{\Delta}
+C\right), 
\label{C}
\eeq
$$
C\approx 3\,\, {\rm at }\,\, \Delta /M_\pi\approx 1.4 .
$$

Parametrically the right-hand side of eq. (\ref{C}) is proportional
to $1/m_{b,c}^2$ (through $\Delta^2$), as it should,
of course. Moreover, the $D\pi$ inelastic contribution is
additionally suppressed by $1/N_c$ (through $\lambda^2$) where
$N_c$ is the number of colors.

Substituting the existing upper bound for $\lambda^2$
\cite{wiseb} ($\lambda^2<0.5f_\pi^{-2}$) as the
actual value of $\lambda^2$ and $\mu_{hadr}
\sim 1$ GeV we find that the
inelastic contribution in the sum rule (\ref{SR!}) due to 
the channel $D$ plus soft pion is close to 7\%.  Approximately one 
third comes
from the logarithmic term and two thirds from the constant. 

Thus, if $\lambda^2$ is close to its upper bound
this inelastic channel alone produces the same effect on $F_{B\ra D^* 
}$ as the $\mu_G^2$  term in 
eq. (\ref{SR!!}), i.e. decreases $F_{B\ra D^* }$
by 0.035.  Thus, we conclude that it is perfectly reasonable and safe
to assume  the inelastic effect to vary between
zero and 100\% of the non-perturbative terms in 
eq. (\ref{SR!!}), and our educated guess is fully substantiated.

Comparing our result with the renormalization of $F_{B\ra D^*}$
due to soft pion exchange considered in \cite{randall}
we observe, with satisfaction, that the elastic renormalization
of Ref. \cite{randall} is exactly the same (parametrically)
and has the opposite sign, so that the combined effect of
the soft pions in the integral (\ref{inth}) vanishes, and this
integral remains equal to unity, as it should be if we limit
ourselves to the infrared contributions and thus neglect the 
non-perturbative corrections in eq. (\ref{SR!}). Numerically our
estimate of eq. (\ref{C})  is  higher, by a factor of $\sim$2, since in 
\re{randall}
$1/M_B$ was set equal to zero and $\Delta$ was approximated
by $M_{D^*}-M_D$. 

\vspace*{.2cm}

4. The CKM matrix element  $|V_{cb}|$ can be alternatively 
determined from 
the inclusive 
semileptonic width $\Gamma{B\ra X_c l\nu}$. The theoretical 
expression for 
the widths is
known in the literature including the $\alpha_s$
and the leading non-perturbative correction,
$$
\Gamma(B\ra l \nu 
X_c)=\frac{G_F^2m_b^5}{192\pi^3}|V_{cb}|^2\cdot \left\{
\left(z_0(x)-\frac{2\al_s}{3\pi}(\pi^2-25/4) z_0^{(1)}(x)\right)\cdot 
\right.$$
\beq
\left. \cdot 
\left(1-\frac{\mu_\pi^2-\mu_G^2}{2m_b^2}\right) -
z_1(x) \frac{\mu_G^2}{m_b^2}
+{\cal O}(\al_s^2, \al_s/m_b^2, 1/m_b^3)\right\} 
\label{slwidth}
\eeq
where the phase space
factors $z$ account for the mass of the final quark,
$$z_0(x)=1-8x+8x^3-x^4-12x^2\log{x} \;\;,\;\; z_1(x)=(1-x)^4\;\;\;, $$
\beq
z_0^{(1)}(0)=1\;\;,\;\;  z_0^{(1)}(1)= 3/(2\pi^2-25/2))\approx 0.41
\;\;\;,\;\;\;x=(m_c/m_b)^2 . 
\label{zs}
\eeq
The function $z_0^{(1)}$ can be found in Refs. \cite{rc}. 
The nonperturbative terms proportional to $\mu_G^2$
and $\mu_\pi^2$ were first found in Ref. \cite{bigi}; all corrections 
together 
are compiled in Refs.\cite{Ds,LS}.

The explicit form 
of the perturbative correction here refers to the one-loop value of 
the 
so called pole 
mass, see e.g. \cite{novikov}. Theoretically this object is ill-defined
\cite{pole}. Generally speaking, if the result for the decay
rate is expressed in terms of the pole mass one should expect
large (factorially divergent) coefficients in the $\as$ expansion.
If nonperturbative terms are included it is necessary to
express the result in terms of an euclidean mass normalized
appropriately \footnote{It was shown in \re{pole} that the $b$ 
quark mass 
that enters the expressions for inclusive widths at the level of 
nonperturbative corrections is a well-defined running mass 
normalized at sufficiently high momentum scale that {\em per se} 
does not 
have any intrinsic theoretical uncertainty. It can be determined to
a very 
good accuracy from the spectra of quarkonia.} 
. Transition from the pole mass to the euclidean mass
might  change the coefficient of the $\alpha_s$ correction.
We do not need to do that, however. Indeed, our point is that
the decay rate (\ref{slwidth}) depends essentially on the difference
of the quark masses, $m_b-m_c$, and in this difference all
uncertainties and definition dependence cancel.  The residual weak 
dependence on
the individual masses is reflected in the error bars we will
ascribe  to our result.

For the pole mass of the $b$ quark (or for the mass
normalized not far from the would-be mass shell) it is reasonable to 
accept
\beq
m_b = 4.8\pm 0.1 \,\, {\rm GeV} .
\label{mb}
\eeq
The central value, 4.8 GeV, follows from the QCD sum rule analysis
of the $\Upsilon$ system \cite{voloshin2}. To be on a safe side we
multiplied the original error bars by a factor of 4.  It is worth
noting that it is very difficult -- practically impossible --
to go outside the indicated limits given the constraint on
$m_b-m_c$ to be discussed below. Indeed the central value of
$m_b$ above implies $m_c \approx 1.30$ GeV (provided
we accept the estimate (\ref{bra}) for $\mu_\pi^2$) which matches
very well with an independent determination of the $c$ quark
pole mass \cite{novikov}. Plus or minus 100 MeV in $m_b$
are translated in $\pm 100$ MeV in $m_c$. It seems perfectly
safe to say that $m_c$ lies between 1.20 and 1.40 GeV --
one can hardly imagine that the one-loop pole $c$ quark mass
is less than $1.20$ or larger than $1.40$. Thus, we believe
that allowing $m_b$ to vary in the interval (\ref{mb})
we fully cover the existing uncertainty in this parameter. We will
{\em not} allow $m_c$ to change independently, however;
this parameter will be tied up to $m_b$. This simple
step dramatically reduces the uncertainty in the theoretical
prediction for $\Gamma (B\ra X_cl\nu)$.

At first sight it might seem that the fifth power of the 
$b$ quark mass in \eq{slwidth} strongly magnifies the uncertainty in 
$m_b$. 
It is possible to check, however, that the  perturbative  expression 
for the 
width, $m_b^5\cdot 
z_0(m_c^2/m_b^2)$,  being  a rather sophisticated
function of both masses, $m_b$ and $m_c$, is sensitive mostly
to the quark mass difference. Moreover, even though
each mass individually has a noticeable scatter
around the canonical central values ($\pm 2\%$ for $b$ and $\pm 
6\%$
for $c$), the quark mass
difference is known to a much better accuracy within HQET
\cite{HQET,HQETr}:
\beq
 m_b-m_c\,= \, \frac{M_{B}+3M_{B^*}}{4} - \frac{M_{D}+3M_{D^*}}{4} 
\,+ \,
\mu_\pi^2 \cdot (\frac{1}{2m_c}-\frac{1}{2m_b}) \,+\, {\cal 
O}(1/m_c^3\,,\;
1/m_b^3) \;\;.
\label{mdif}
\eeq
This relation holds for the masses normalized at the scale 
below the mass of the $c$ quark. However, the infrared renormalon 
singularities cancel in this difference, and it is free of intrinsic 
theoretical uncertainties at any normalization point.

The expression for the width has
such a structure that for the given fixed value of $m_b-m_c$
the variation of the individual masses inside the allowed
interval affects the theoretical prediction at a much
weaker level than what one gets in the standard approach,
with both masses changing independently. This observation noted 
long ago 
in
phenomenological studies has a reason:  in the 
semileptonic decays in a large part of the
phase space  effectively we are not far from the 
SV limit, i.e. $|\vec q|<M_D$.  
Exactly in this limit, at $|\vec q|/M_D\ra 0$, the
semileptonic transition would depend {\em only}
on the quark mass difference. The most clear-cut
manifestation of the proximity to the SV limit is the fact that
about 60\% of the semileptonic decay rate is due to the
``elastic" channels, $B\ra D l\nu$ and $B\ra D^* l\nu$; in the SV limit
these two channels completely saturate the probability (up to
non-perturbative corrections of order $\Lam^2/m_{b,c}^2$). 
This reason explains why we expect the 
higher order perturbative corrections to behave 
in the same manner.

In this way we get numerically
\beq
|V_{cb}| = 
0.0415\left(\frac{1.49\mbox{ps}}{\tau_B}\right)^{1/2}\left( 
\frac{BR_{sl}(B)}{0.106}\right)^{1/2}
\label{number} 
\eeq
where we used the central value $4.80\GeV$ (\re{voloshin2}) 
for the one-loop pole mass of the $b$ quark, and the value of the 
strong 
coupling $\as=0.22$;  the expectation value
of the kinetic energy is also set equal to its central value, 
$\mu_\pi^2=0.54\GeV^2$.

Let us discuss what error bars in eq. (\ref{number}) theory is
responsible for.  First,
 the variation of $m_b$ (or, 
alternatively, $m_c$) in the range $\pm 100\MeV$ results only in a 
$\mp 
1.6\%$ relative variation of $|V_{cb}|$ if other parameters are kept 
fixed! 
The most sizable uncertainty arises in this approach due to dependence 
of $m_b-m_c$ on the value of 
$\mu_\pi^2$. Again, to be on a safe side, we double
the original theoretical error bars \cite{Braun} in this parameter and 
allow it 
to vary within the following limits: 
\beq
0.35\GeV^2\lsim \mu_\pi^2 \lsim 0.8\GeV^2 .
\label{kinetic}
\eeq
This uncertainty leads to the  change in $|V_{cb}|$ of $\mp 2.8\%$, 
with 
practically linear
dependence. It seems obvious  that 
the interval (\ref{kinetic})   overestimates the existing uncertainty in
$\mu_\pi^2$. 

It is worth noting that the value of $\mu_\pi^2$ can, and will be 
measured 
soon via the shape 
of the lepton spectrum in $b\ra c l\nu$ inclusive decays 
\cite{optical} with 
theoretical accuracy of at least $\lsim 0.1\GeV^2$.

Finally there is some dependence on the value of the strong coupling 
in 
eq.~(\ref{slwidth}). Numerically the uncertainty constitutes about 
$\pm 1\%$ when 
$\as$ is 
varied 
between $0.2$ and $0.25$. This must and will be reduced by explicit 
calculation of 
the next loop correction, which is straightforward (though somewhat 
tedious 
in 
practice). All perturbative QCD corrections are 
well behaved here: no logarithms of the mass ratio can appear, at 
least if 
the result is expressed in terms of the quark masses and $\as$ 
normalized at a proper euclidean scale.

Therefore, the above numerical estimates imply that already at 
present the theoretical uncertainty in the ``inclusive" value of 
$|V_{cb}|$ does not exceed $\sim\pm 5\%$ and is quite competitive 
with
the existing experimental uncertainties in this quantity. It seems 
possible to
further reduce this error to 4 or even 3\%
by measuring $\mu_\pi^2$ and calculating the 
two-loop perturbative correction to the width. The inclusive method,
thus, is not 
only  more  accurate experimentally (due to higher statistics than in 
the 
exclusive case)  but  it also seems more promising 
from the side of both, the  theoretical uncertainties as they exist now 
and 
their  future possible reduction.

\vspace*{.2cm}

5. Let us return now to the exclusive method. 
Using the value of $|V_{cb}|$ determined from the inclusive 
semileptonic 
width and our  estimates of the $F_{B\ra D^*}$, \eq{prefer},  one can 
determine the position of the theoretical prediction for the 
extrapolation of 
the exclusive decay rate to  zero recoil. 
Alternatively, one can express the 
same result by explicitly introducing  the correction 
due to the fact that $F_{B\ra D^*}\neq 1$
in  eq.~(\ref{exp}).  Assuming $F_{B\ra D^*}=0.89$ and 
using the value from eq.~(\ref{exp}) one obtains 
\beq
|V_{cb}| = 0.043\pm 0.007\; . 
\label{corr}
\eeq
The agreement with the ``inclusive'' result (\ref{number}) seems to be 
even 
too good taking into consideration the experimental uncertainties and 
the 
theoretical 
assumptions involved in the estimates of $F_{B\ra D^*}$.

\vspace*{.2cm}

6. It is instructive to make a brief comparative analysis of  the 
theoretical 
uncertainties in  the two 
methods: inclusive versus exclusive. In the both cases 
there 
are no nonperturbative corrections at the level $1/m_{b,c}$. The 
main 
difference,  however, 
is that the leading $1/m_{b,c}^2$ nonperturbative corrections {\em 
can} be, 
and 
have been fully  calculated for the inclusive widths, whereas they 
{\em can not}
be determined in a model-independent way for the exclusive $B\ra 
D^{*}$ 
form factor even at zero recoil (see e.g. \cite{mannel}).  Either we 
have a
contamination due to the contribution from higher states (in our 
approach),
or non-local contributions within  the more traditional approach
of Refs. \cite{FN,mannel}. The latter are poorly controllable. 
It is fortunate that in our approach the $\mu_G^2$ and $\mu_\pi^2$ 
terms
make $F_{B\ra D^*}$ smaller than unity; the contamination due to the 
higher
states, being positive-definite, works in the same direction. 

The exclusive approach has 
certain conceptual advantages: apart from the form factor itself, the 
measured rate is 
given in terms of the masses of real $B$ and $D^*$ mesons, whereas 
in 
calculating the inclusive widths one uses -- though well defined 
theoretically, 
but still uncertain at some level -- quark masses. 
On the other hand, the total semileptonic 
width, in turn, has a theoretical advantage over the exclusive 
predictions: in 
the inclusive approach  determination of the CKM matrix elements 
is  meaningful  even in the limit the  light final quark (i.e. for $b\ra 
u$
transition), 
whereas the exclusive approach becomes useless due to our 
ignorance
of the heavy-to-light form factors even at zero recoil.  (In the 
opposite limit when $m_c$ increases approaching $m_b$ -- the 
gold-plated 
situation for exclusive form factors -- the OPE-based analysis of the 
inclusive transitions automatically reproduces all  results of  HQET 
\cite{optical}.)

The perturbative corrections have been calculated in both cases to 
one loop. 
The one-loop gluon correction turns out to be somewhat smaller for 
the 
exclusive 
transition 
due to numerical cancellations. Although some 
higher order logarithmic summation has been performed in the 
exclusive 
case, 
it does not seem to be very useful for $m_b/m_c\sim 3$. Logarithmic 
terms
of this type 
do not appear at all in the  inclusive widths  due to 
the infrared stability of the latter. 
The real improvement of the perturbative estimates 
can be achieved only by the exact two-loop calculation of the $\as^2$ 
terms in 
both cases.

Summarizing, numerically
and statistically the inclusive approach seems to win.

\vspace*{1cm}

{\large\bf Acknowledgments} \hspace*{.7cm} 
This work has been done during our stay at Institute of Theoretical 
Physics,
University of California (Santa Barbara). We are grateful to L. Randall
and D. Kaplan for inviting us to participate in the program on heavy
quarks at ITP. We would like to thank our colleagues and the staff
of ITP for kind hospitality. Useful discussions with I. Bigi, 
A. Falk, Y. Kubota, R. Poling, L. Randall, M. Voloshin and M. Wise are  
acknowledged. We are grateful to L. Randall for pointing out 
Ref. \cite{randall} to us. 

The results of this paper were 
reported by N.U. at the 1994 APS meeting, Crystal City, VA, April 
18-22. This work was supported in part by DOE under the
grant number DE-FG02-94ER40823 and by NSF under the grant
number NSF-PHY89-04035.

\newpage

\begin{center}
{\bf Figure Captions}
\end{center}

The graps determining the amplitude $\langle D\pi |A_i |B\rangle$
in the soft-pion limit.

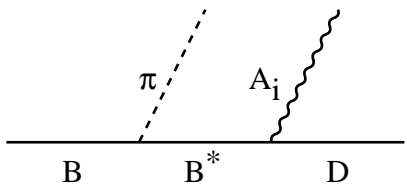
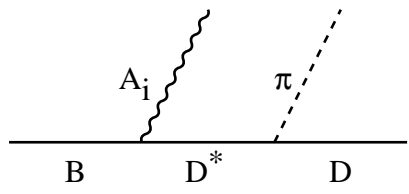
\end{document}